\documentclass[preprint,3p,twocolumn]{elsarticle}

\usepackage{epsfig}
\usepackage{amssymb}
\usepackage{psfrag}
\usepackage{rotating}
\usepackage{multirow}
\usepackage{url}
\usepackage[latin1]{inputenc}

\journal{International Journal of Hydrogen Energy}
\newcommand\T{\rule{0pt}{2.6ex}}

\newcommand{\psfragall}{\psfrag{d}{\small{Davis}} \psfrag{g}{\small{GRI-Mech}} \psfrag{k}{\small{Konnov}}  \psfrag{l}{\small{Leeds}}  \psfrag{i}{\small{Li}}
\psfrag{o}{\small{\'{O} Conaire}}  \psfrag{s}{\small{SanDiego}}}

\newcommand{\psfragxyign}{\psfrag{y}{\small{$\tau_{ign}$ [$\mu$s]}}  \psfrag{x}{\small{$P$ [atm]}}}

\newcommand{\psfragxysl}{\psfrag{y}{\small{$S_L$ [cm/s]}}

\psfrag{x}{\small{$P$ [atm]}}}

\begin{document}

\begin{frontmatter}

\title{Assessment of existing H$_2$/O$_2$ chemical reaction mechanisms at reheat gas turbine conditions }

\author[sintef]{Torleif Weydahl \corref{cor1}}

\author[alstom]{Madhavan Poyyapakkam}

\author[sintef]{Morten Seljeskog}

\author[sintef]{Nils Erland L. Haugen}

\cortext[cor1]{Corresponding author Torleif.Weydahl@sintef.no}

\address[sintef]{SINTEF Energy Research, N-7465; Trondheim, Norway}
\address[alstom]{Alstom, Switzerland}

\begin{abstract}
This paper provides detailed comparisons of chemical reaction mechanisms of H$_2$ applicable at high preheat temperatures and pressures relevant to gas turbine and particularly Alstom's reheat gas turbine conditions. It is shown that the available reaction mechanisms exhibit large differences in several important elementary reaction coefficients. The reaction mechanisms are assessed by comparing ignition delay and laminar flame speed results obtained from CHEMKIN with available data, however, the amount of data at these conditions is scarce and a recommended candidate among the mechanisms can presently not be selected. Generally, the results with the GRI-Mech and Leeds mechanisms deviate from the Davis, Li, \'{O} Conaire, Konnov and San Diego mechanisms, but there are also significant deviations between the latter five mechanisms that altogether are better adapted to hydrogen. The differences in ignition delay times between the dedicated hydrogen mechanisms (\'{O} Conaire, Li and Konnov) range from approximately a maximum factor of 2 for the H$_2$-air cases, to more than a factor 5 for the H$_2$/O$_2$/AR cases. The application of the computed ignition delay time to reheat burner development is briefly discussed.

\end{abstract}

\end{frontmatter}

\section{Introduction}
\label{sec:intro}
Among the technologies for fossil fuel power production with CO$_{2}$
capture, \textquotedblleft pre-combustion\textquotedblright\ removal of CO$_{2}$ is one of the strong candidates. Pre-combustion technology for stationary power production implies burning hydrogen or hydrogen rich mixtures in a gas turbine. Utilizing hydrogen in non-premixed burners without dilution (steam or nitrogen) causes unacceptable levels of NO$_{x}$ emissions due to the high flame temperature \cite{chiesa2005}. To avoid the higher cost and maintenance that comes with dilution, there is an increasing interest in developing non-diluted premixed or partially premixed hydrogen burners. 

\begin{figure}[tbp]
\centering
\includegraphics[width=7.5cm]{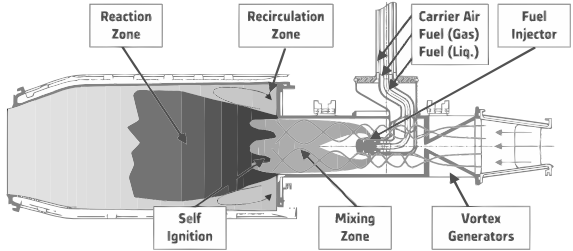} 
\caption{Operation principle of Sequential EnVironmental (SEV) combustor. Illustration from Ciani et al.~\cite{ciani2010}}
\label{fig:sev}
\end{figure}

Hydrogen is a particular molecule with high diffusivity and reactivity and there are challenges to safe and stable operation of burners utilizing non-diluted premixed hydrogen. One combustion technology where premix combustion of hydrogen has a potential is the reheat or sequential gas turbine technology. Other related applications are ramjet, scramjet and afterburner technologies. The reheat gas turbine technology was first commercialized in 1948 by Brown Boveri Co., and is marketed today by Alstom. These turbines had the sequential combustor design, which is used to increase efficiency and provide operational flexibility while managing low emissions. The combustion system uses an EnVironmental (EV) burner in the first combustion stage followed by a Sequential EV (SEV) burner, illustrated in Figure \ref{fig:sev}, in the second stage \cite{ciani2010,doebbeling2005}. Since the reheat combustor is fed by high-temperature expanded exhaust gas of the first combustor, the operating conditions allow auto-ignition (spontaneous ignition) of the fuel air mixture without additional energy being supplied to the mixture. To prevent ignition of the fuel air mixture in the mixing region, an appropriate distribution of the fuel across the burner exit area must be obtained and the residence time in the mixing region must not exceed the auto-ignition delay time. As of today with conventional fuels, this is solved by using delta wing shaped vortex generators to mix the fuel, which is injected in a centrally positioned lance \cite{eroglu2001}.

The chemical mechanism is an important element in several tools for gas turbine calculations. Computational Fluid Dynamics (CFD) has evolved to become one of the main tools in design of gas turbine combustors, especially in the developing stage where experiments are relatively resource intensive and expensive. Accurate CFD predictions of hydrogen combustion rely on high precision in the chemical mechanisms.

A number of detailed mechanisms for hydrogen, often known as the core of any detailed hydrocarbon combustion reaction mechanism \cite{konnov2008}, have been developed during the last two to three decades. Some of these are sub-mechanisms of larger hydrocarbon mechanisms such as methane or ethane. A brief overview of the mechanisms considered in the present work is given in the following and summarized in Table \ref{tab:overview}. The H$_2$/O$_2$ mechanism of Li et al.~\cite{li2004} is based on the mechanism that originally was developed by F.L.~Dryer at Princeton University and further developed by Mueller et al.~\cite{mueller1999}. This mechanism has been compared against a wide range of experimental conditions with temperatures between 298-3000\,K, pressures from 0.3-87\,atm and equivalence ratios from 0.25-5. The mechanism of Mueller was also used as the basis for the mechanism developed by \'{O} Conarie et al.~\cite{conaire2004} at University of Ireland in Galway. The series of experiments numerically investigated ranged from 298-2700\,K, pressures between 0.05-87\,atm and equivalence ratios from 0.2-6. Konnov~\cite{konnov2008} has derived a H$_2$/O$_2$ mechanism from a methane mechanism  with a recent review and update of the elementary reactions. This mechanism is validated with ignition experiments (950 to 2700\,K, sub-atmospheric to 87\,atm) and flame speeds at pressures from 0.35-4\,atm. Davis et al.~\cite{davis2005} developed a H$_2$/CO/O$_2$ mechanism based on an extensive validation against ignition delay and flame speed data in the range from 0.05-64\,atm and 298-1754\,K (for hydrogen). The San Diego mechanism developed by Forman Williams and co-workers at the University of California, San Diego \cite{williams2007,petrova2006} is also studied here. The full mechanism includes hydrocarbons up to C3 \cite{williamsweb}. The H$_2$/O$_2$ part has been validated against hydrogen ignition delay data up to 33\,bar \cite{delalamo2004}. These five mechanisms have been selected for consideration based on thoroughness in validation against hydrogen ignition and flame speed data. In addition we consider two of the most widely used hydrocarbon mechanism that includes a H$_2$/O$_2$ subset. The Leeds methane mechanism has been developed at Leeds University by Hughes et al.~\cite{hughes2001}, while the GRI mechanism origins from the Gas Research Institute \cite{grimech}. More comprehensive summaries of the various mechanisms for hydrogen combustion are for example given by del Alamo et al.~\cite{delalamo2004} and Ströhle and Myhrvold \cite{strohle2007}.

\begin{table}[htp]
\centering
\begin{tabular}{l|l|l|l|l}
\rule{0pt}{2.6ex} Mechanism & $N_s$ & $N_{R}$ & $N_R^{'}$ & Reference \\ \hline\hline
\rule{0pt}{2.6ex} Li & 10 & 23 & 23   & \cite{li2004} \\
\rule{0pt}{2.6ex} \'{O} Conaire & 10 & 21 & 21 & \cite{conaire2004}/\cite{conaireweb} \\
\rule{0pt}{2.6ex} Konnov & 10 & 33 & 33 & \cite{konnov2008} \\  
\rule{0pt}{2.6ex} Davis & 14 & 38 & 25 & \cite{davis2005}/\cite{davisweb} \\ 
\rule{0pt}{2.6ex} San Diego & 46 & 235 & 21 & \cite{williams2007}/\cite{williamsweb} \\ 
\rule{0pt}{2.6ex} Leeds & 37 & 175 & 23 &\cite{hughes2001}/\cite{leedsweb} \\
\rule{0pt}{2.6ex} GRI-Mech & 53 & 325 & 28 & \cite{grimech} \\ 
\hline\hline
\end{tabular}%
\caption{Overview of the chemical kinetic mechanisms and their most recent references considered in the present work. $N_s$  is the total number of species in the mechanism (including Argon). $N_R$ is the total number of reactions in the mechanism while $N_R^{'}$ is the number of reactions
in the H$_2$/O$_2$ subset (duplicate reactions and separately formulated third body reactions are counted).} 
\label{tab:overview}
\end{table}

The number of available mechanisms and validation studies are significant, and still there are uncertainties related to the elementary reactions involved \cite{konnov2008}. Very recently, Burke et al.~\cite{burke2010} investigated the mass burning rate (the product of laminar flame speed and unburnt gas density) and found that there is a negative pressure dependence at high pressures and low temperatures. According to their results, this dependence is not captured by the available mechanisms due to uncertain and even missing third-body reactions.

The aim of the present study is to investigate the performance of the selected detailed chemical mechanisms at the high temperature and pressure conditions relevant to gas turbine and particularly reheat engine conditions. As a starting point in reheat burner design, the burner residence time is chosen by considering the ignition times. This is based on the assumption that the flame position is controlled by pure auto-ignition. The effect of pressure on ignition delay time and laminar flame speed is investigated in order to illustrate the effect of variable engine load. For the universality of this study, pressures also higher than the baseload operating pressure of the Alstom SEV burner are considered. The data of Herzler and Naumann \cite{herzler09} and Petersen et al.~\cite{petersen1996} are applied for comparison. There is generally little data available for auto-ignition and laminar flame speed in the pressure range from 15 to 30\,bar particularly at high temperatures. Consequently the mechanisms are also compared at conditions where they have not been validated against experiments. Nevertheless, for predictions of ignition delay and flame speed under reheat gas turbine conditions, their performance at these conditions is of interest.

\section{Overview of important elementary reactions}
\label{sec:overview}
 In Table~\ref{tab:reacrates} the most important H$_2$/O$_2$ reactions for ignition, extinction and flame propagation identified by Ströhle and Myhrvold \cite{strohle2006} are listed. We have also verified that for the results in the present work this 11-step reduced Li mechanism closely reproduces the results with the full Li mechanism. This has, however, not been verified for the other selected mechanisms. The reactions coefficients marked with ${(*)}$ are for the reverse reaction and can not be directly compared. Among the selected reactions, the reverse reaction (10) is the main initiating step which creates H-radicals. The chain branching/propagating steps (1-3) are important for the auto-ignition, but competes with the chain terminating steps (9a-f) that suppress the formation of the highly reactive H-radical \cite{williams2007,delalamo2004}. The chain terminating steps are pressure dependent and become increasingly important at higher pressures.

In the following, a brief overview of the differences between the reaction coefficients listed in Table~\ref{tab:reacrates} and \ref{tab:additionals} are given. More comprehensive and in-dept discussions of these coefficients are found elsewhere in the literature \cite{konnov2008,baulch1994,baulch2005}. The chain-branching reaction (1) is recognized as one of the most sensitive and important reactions to hydrogen combustion, and the differences between the reaction coefficients are also significant. For reference, Konnov \cite{konnov2008} applies the latest expression recommended by Baulch et al.~\cite{baulch2005}. The coefficients in the reaction (2-3) show little deviation for all mechanisms. Li et al.~\cite{li2004} modified the pre-exponential factor of reaction (8a) to improve flame speed predictions. Apart from this, the mechanisms deviate mainly in the third-body collision efficiencies. It should for instance be noted that the third-body effect of H$_2$O in Davis is about half the quantity in the Li, \'{O} Conarie and San Diego mechanisms. Konnov applies a separate equation (8b) listed in Table~\ref{tab:additionals} when the collision partner is H$_2$O. This is effectively to take temperature-dependent third-body effects into account \cite{delalamo2004}. For the chain terminating reaction (9a), the GRI-Mech and Leeds mechanisms do not include low-pressure limits. In addition, GRI-Mech and Leeds have separate expressions for the collision partners H$_2$/H$_2$O and H$_2$O, respectively.  Also Konnov applies separate fall-off expressions for individual collision partners. Among the other mechanisms, the differences in the low-pressure limit coefficients and in collision efficiencies especially for H$_2$ and H$_2$O are the most apparent differences. The Li-mechanism has an alternative reaction (9a) when the main bath is AR where the low pressure limit coefficients are [9.0$\cdot10^{19}$, $-1.5$, 492], the broadening factor is F$_c$=0.5, and the third body efficiencies are O$_2$=1.1, H$_2$O=16 and H$_2$=3. The reverse reaction (10) is the most important initiating step in the mechanisms \cite{williams2007}. The differences between Konnov, which uses the coefficients recommended by Baulch et al.~\cite{baulch2005}, and Li, \'{O} Conarie and San Diego should be noted. Konnov \cite{konnov2008} reports large uncertainty of the HO$_2$ reaction (13). The difference between the mechanisms is evident in that additional duplicate reactions are introduced in the GRI-Mech, Davis and Konnov mechanisms. However, the coefficients in the respective other mechanisms are close to equal for this reaction. For reference, Li uses the expression recommended by Baulch et al.~\cite{baulch1994,baulch2005} for reaction (14). All mechanisms except Davis use duplicate reactions with similar coefficients for reaction (14) and (14dup). Considering reaction (17) the mechanisms exhibit large variations. Again, Li, \'{O} Conarie and San Diego have the closest agreement, while Konnov, which adopts the Baulch et al.~\cite{baulch2005} recommendations, has lower pre-exponential factor and activation energy. The expressions applied in GRI-Mech and Davis involve temperature dependences of the pre-exponential factor.

\begin{table*}[htp]
  \centering
 \footnotesize{
  \begin{tabular}{p{2.9cm}|p{1.35cm}|p{1.35cm}|p{1.35cm}|p{1.35cm}|p{1.35cm}|p{1.35cm}|p{1.35cm}}
  
    Reaction & GRI-Mech & Davis & Konnov & Li  & \'{O} Conaire & San Diego & Leeds \\ \hline \hline
(1)~H+O$_2$ = OH+O &
\T 2.65$\cdot10^{16}$ $-0.7$ 17041 &
\T 2.64$\cdot10^{16}$ $-0.7$ 17041 &
\T 2.06$\cdot10^{14}$ $-0.1$ 15022 &
\T 3.55$\cdot10^{15}$ $-0.4$ 16599 &
\T 1.91$\cdot10^{14}$ 0.0  16440 &
\T 3.52$\cdot10^{16}$ $-0.7$ 17070 &
\T 9.76$\cdot10^{13}$ 0.0  14821
 \\ \hline
(2)~O+H$_2$ = OH+H &
\T 3.87$\cdot10^{4}$ 2.7 6260 &
\T 4.59$\cdot10^{4}$ 2.7 6260 &
\T 5.06$\cdot10^{4}$ 2.7 6290 &
\T 5.08$\cdot10^{4}$ 2.7 6290 &
\T 5.08$\cdot10^{4}$ 2.7 6292 &
\T 5.06$\cdot10^{4}$ 2.7 6291 &
\T 5.12$\cdot10^{4}$ 2.7 6277
 \\ \hline
(3)~H$_2$+OH = H$_2$O+H &
\T 2.16$\cdot10^{8}$ 1.5 3430 &
\T 1.73$\cdot10^{8}$ 1.5 3430 &
\T 2.14$\cdot10^{8}$ 1.5 3450 &
\T 2.16$\cdot10^{8}$ 1.5 3430 &
\T 2.16$\cdot10^{8}$ 1.5 3430 &
\T 1.17$\cdot10^{9}$ 1.3 3635 &
\T 4.52$\cdot10^{8}$$^{(*)}$ 1.6 18401
 \\ \hline \hline
(8a) H+OH+M = H$_2$O+M &
\T 2.20$\cdot10^{22}$ $-2.0$ 0.0 H$_2$O=3.65 H$_2$=0.73 AR=0.38 &
\T 4.40$\cdot10^{22}$ $-2.0$ 0.0 H$_2$O=6.3 H$_2$=2.0 AR=0.38 &
\T 6.06$\cdot10^{27}$$^{(*)}$ $-3.3$ 120770 H$_2$O=0 H$_2$=3.0 N$_2$=2.0 O$_2$=1.5 &
\T 3.80$\cdot10^{22}$ $-2.0$ 0.0 H$_2$O=12 H$_2$=2.5 AR=0.38 &
\T 4.50$\cdot10^{22}$ $-2.0$ 0.0 H$_2$O=12 H$_2$=0.73 AR=0.38 &
\T 4.00$\cdot10^{22}$ $-2.0$ 0.0 H$_2$O=12 H$_2$=2.5 AR=0.38 &
\T 5.53$\cdot10^{22}$ $-2.0$ 0.0 H$_2$O=2.55 N$_2$=0.4 O$_2$=0.4 AR=0.15
 \\ \hline

(9a) H+O$_2$(+M) = HO$_2$(+M) &
\T 2.80$\cdot10^{18}$ $-0.9$ 0.0 O$_2$=0 H$_2$0=0 N$_2$=0 AR=0&
\T 5.12$\cdot10^{12}$ 0.4 0.0 [6.3$\cdot10^{19}$ $-1.4$ 0.0] F$_c$=0.5 O$_2$=0.85 H$_2$0=12 H$_2$=0.75 AR=0.4 &
\T 4.66$\cdot10^{12}$ 0.4 0.0 [5.7$\cdot10^{19}$ $-1.4$ 0.0] F$_c$=0.5 O$_2$=0 H$_2$0=0 H$_2$=1.5 AR=0 &
\T 1.48$\cdot10^{12}$ 0.6 0.0 [6.4$\cdot10^{20}$ $-0.17$ 5248] F$_c$=0.8 O$_2$=0.78 H$_2$0=11 H$_2$=2  &
\T 1.48$\cdot10^{12}$ 0.6 0.0 [3.5$\cdot10^{17}$ $-0.4$ $-1120$] F$_c$=0.5 O$_2$=0.78 H$_2$0=14 H$_2$=1.3  &
\T 4.65$\cdot10^{12}$ 0.4 0.0 [5.7$\cdot10^{19}$ $-1.4$ 0.0] F$_c$=0.5 H$_2$0=16 H$_2$=2.5 AR=0.7  &
\T 2.10$\cdot10^{18}$ $-0.8$ 0.0 O$_2$=0.4 H$_2$0=0 H$_2$=.75  N$_2$=0.67 AR=0.29 
\\ \hline  \hline   
 (10)~HO$_2$+H = H$_2$+O$_2$ &
\T 4.48$\cdot10^{13}$ 0.0 1068 &
\T 5.92$\cdot10^{5}$$^{(*)}$ 2.4 53502 &
\T 1.05$\cdot10^{14}$ 0.0 2047 &
\T 1.66$\cdot10^{13}$ 0.0 823 &
\T 1.66$\cdot10^{13}$ 0.0 820 &
\T 1.66$\cdot10^{13}$ 0.0 823 &
\T 4.28$\cdot10^{13}$ 0.0 1408
 \\ \hline
 (11) HO$_2$+H = OH+OH &
\T 8.40$\cdot10^{13}$ 0.0 635.0 &
\T 7.48$\cdot10^{13}$ 0.0 295.0 &
\T 1.90$\cdot10^{14}$ 0.0 875.0 &
\T 7.08$\cdot10^{13}$ 0.0 295.0 &
\T 7.08$\cdot10^{13}$ 0.0 300.0 &
\T 7.08$\cdot10^{13}$ 0.0 295.0 &
\T 1.69$\cdot10^{14}$ 0.0 883.0
 \\ \hline
(13) HO$_2$+OH = H$_2$O+O$_2$ &
\T 1.45$\cdot10^{13}$ 0.0 $-500$ &
\T 2.38$\cdot10^{13}$ 0.0 $-500$ &
\T 2.89$\cdot10^{13}$ 0.0 $-500$ &
\T 2.89$\cdot10^{13}$ 0.0 $-497$ &
\T 2.89$\cdot10^{13}$ 0.0 $-500$ &
\T 2.89$\cdot10^{13}$ 0.0 $-497$ &
\T 2.89$\cdot10^{13}$ 0.0 $-500$
 \\ \hline
(13dup) HO$_2$+OH = H$_2$O+O$_2$ &
\T 5.00$\cdot10^{15}$ 0.0 17330 &
\T 1.00$\cdot10^{16}$ 0.0 17330 &
\T 9.27$\cdot10^{15}$ 0.0 17500
&
&
&
 \\ \hline 
(14) HO$_2$+HO$_2$ = H$_2$O$_2$+O$_2$ &
\T 4.20$\cdot10^{14}$ 0.0 12000 &
\T 3.66$\cdot10^{14}$ 0.0 12000 &
\T 1.03$\cdot10^{14}$ 0.0 11040 &
\T 4.20$\cdot10^{14}$ 0.0 11982 &
\T 4.20$\cdot10^{14}$ 0.0 11980 &
\T 3.02$\cdot10^{12}$ 0.0 1386.2 &
\T 4.22$\cdot10^{14}$ 0.0 11957
 \\ \hline
(14dup) HO$_2$+HO$_2$ = H$_2$O$_2$+O$_2$ &
\T 1.30$\cdot10^{11}$ 0.0 $-1630$ &
\T 1.30$\cdot10^{11}$ 0.0 $-1630$ &
\T 1.94$\cdot10^{11}$ 0.0 $-1409$ &
\T 1.30$\cdot10^{11}$ 0.0 $-1629$ &
\T 1.30$\cdot10^{11}$ 0.0 $-1629$ &
&
\T 1.32$\cdot10^{11}$ 0.0 $-1623$
 \\ \hline
(15a) OH+OH(+M) = H$_2$O$_2$(+M) &
\T 7.40$\cdot10^{13}$ $-0.4$ 0.0 [2.3$\cdot10^{18}$ $-0.9$ $-1700$] H$_2$O=6 H$_2$=2 AR=0.7 &
\T 1.11$\cdot10^{14}$ $-0.4$ 0.0 [2.0$\cdot10^{17}$ $-0.58$ $-2293$] H$_2$O=6 H$_2$=2 AR=0.7 &
\T 1.00$\cdot10^{14}$ $-0.4$ 0.0 [2.4$\cdot10^{19}$ $-0.8$ 0.0]  F$_c$=0.5 H$_2$O=0&
\T 2.95$\cdot10^{14}$$^{(*)}$ 0.0 48430 [1.2$\cdot10^{17}$ 0.0 45500]  F$_c$=0.5 H$_2$O=12 H$_2$=2.5 AR=0.64 &
\T 2.95$\cdot10^{14}$$^{(*)}$ 0.0 48400 1.3$\cdot10^{17}$ 0.0 45500]  F$_c$=0.5 H$_2$O=12 H$_2$=2.5 AR=0.64&
\T 7.40$\cdot10^{13}$ $-0.4$ 0.0 [2.3$\cdot10^{18}$ $-0.9$ $-1700$] H$_2$O=6 H$_2$=2 AR=0.4 &
\T 7.23$\cdot10^{13}$ $-0.4$ 0.0 [5.6$\cdot10^{19}$ $-0.76$ 0.0] H$_2$O=6.5 N$_2$=0.4 O$_2$=0.4 AR=0.35
 \\ \hline
(17) H$_2$O$_2$+H = HO$_2$+H$_2$ &
\T 1.21$\cdot10^{7}$ 2.0 5200 &
\T 6.05$\cdot10^{6}$ 2.0 5200 &
\T 1.70$\cdot10^{12}$ 0.0 3755 &
\T 4.82$\cdot10^{13}$ 0.0 7950 &
\T 6.03$\cdot10^{13}$ 0.0 7950 &
\T 4.79$\cdot10^{13}$ 0.0 7959 &
\T 1.69$\cdot10^{12}$ 0.0 3747
 \\ \hline \hline
\end{tabular}
}
\caption{\label{tab:reacrates} Overview and comparison of reaction rate coefficients for the most important H$_2$/O$_2$ reactions. For each mechanism and reaction the pre-factor, temperature exponent and activation energy are listed in  respective order. The low-pressure limits of the reactions are given within brackets [ ]. The unit of activation energy is cal/mole. Numbering of the reactions is in accordance with Li et al.~\cite{li2004}. Reactions marked with (*) are reverse reaction coefficients.}
\end{table*}

\begin{table}[htp]
  \centering
   \footnotesize{
  \begin{tabular}{p{2.3cm}|p{1.25cm}|p{1.35cm}|p{1.3cm}}
    Reaction & GRI-Mech & Konnov & Leeds \\ \hline \hline
 (8b) H+OH+H$_2$O = H$_2$O+H$_2$O &
&
\T 1.0$\cdot10^{26}$$^{(*)}$ $-$2.4 120160 
&
 \\ \hline
(9b) H+O$_2$(+O$_2$) = HO$_2$(+O$_2$) &
&
\T 4.7$\cdot10^{12}$ 0.4 0.0  [5.7$\cdot10^{18}$ $-$1.1 0.0] F$_c$=0.5
&
 \\ \hline
(9c) H+O$_2$(+H$_2$O) = HO$_2$(+H$_2$O) &
&
\T 9.1$\cdot10^{12}$ 0.2 0.0  [3.7$\cdot10^{19}$ $-$1.0 0.0] F$_c$=0.8
&
 \\ \hline
 (9c) H+O$_2$(+AR) = HO$_2$(+AR) &
&
\T 4.6$\cdot10^{12}$ 0.4 0.0  [7.4$\cdot10^{18}$ $-$1.2 0.0] F$_c$=0.5
&
 \\ \hline
(9e) H+O$_2$+H$_2$O = HO$_2$+H$_2$O &
\T 1.13$\cdot10^{19}$ $-$0.8 0.0 & &
\T 6.89$\cdot10^{15}$ 0.0 $-$2076
 \\ \hline
(9f) H+O$_2$+N$_2$ = HO$_2$+N$_2$ &
\T 2.6$\cdot10^{19}$ $-$1.2 0.0 &
\\ \hline
 (15b) OH+OH(+H$_2$O) = H$_2$O$_2$(+H$_2$O) &
&
\T 1.0$\cdot10^{14}$ $-$0.37 0.0 [1.45$\cdot10^{18}$ 0.0 0.0] F$_c$=0.5
&
 \\ \hline
\end{tabular}

}
\caption{\label{tab:additionals} Overview and comparison of reaction rates specific to the GRI-Mech, Konnov and Leeds mechanisms.  See the caption of Table~\ref{tab:reacrates} for explanation. }
\end{table}

\section{Method}
\label{sec:method}
Ignition delay time is computed by solving the perfectly stirred closed reactor equations 
\begin{equation}
\frac{dY_i}{d t}=\omega_i, \ \   \frac{d h}{d t}=0 ,\ \ \frac{d p}{d t}=0, 
\end{equation}
where $Y_i$ and $\omega_i$ are mass fraction and chemical reaction rate of species $i$, respectively. The initial value problem is solved at constant enthalpy $h$ (adiabatic) and pressure $p$ conditions using an in-house code where the CHEMKIN library \cite{kee1996} is used for the property and source term calculations. Unless otherwise stated, the ignition delay time is defined by the time when the temperature has reached 200\,K above the initial value. A similar criterion is used by e.g.~Konnov \cite{konnov2008}.

The freely propagating one-dimensional laminar premixed flame calculations were performed using the PREMIX code \cite{kee1985}. The effect of species transport by temperature gradients, the Soret effect, and multicomponent diffusion is included in the computations since these phenomena are of particular importance in hydrogen combustion. The first-order upwind discretization scheme was used with sufficient grid refinement so that a grid-independent solution was achieved for all cases. Since the premixed reactant mixture is above its auto-ignition temperature it is important to ensure that the flame is positioned close enough to the inlet boundary so that the results are not affected by auto-ignition. On the other hand the flame must not be positioned \emph{too} close, since the flame speed calculations are assuming zero gradient on the inlet boundary. Due to these high-temperature related restrictions, the flame speed results were not completely independent of the model input fixed temperature T$_\mathrm{FIX}$.\footnote{When solving the freely propagating flame problem, an additional boundary condition is required. This condition is the temperature T$_\mathrm{FIX}$ of the flame at a given position \cite{kee1985}.} A variation in T$_\mathrm{FIX}$ with $\pm 50$\,K, which generally was chosen close to the inlet temperature value, resulted in a variation in flame speed with 4-5\,\%. Under these high temperature and pressure conditions, achieving experimental results of laminar flame speeds will be extremely challenging. Despite of this the laminar flame speeds have been included in this work because generally in burner design both ignition delay and flame speed are used.

%
%
%
%
\section{Results and Discussion}
\label{sec:result-discussion}

\subsection{Ignition delay results}
\label{sec:igndelay}
\begin{figure}[tbp]
\centering
 \psfragall
 \psfrag{y}{\small{$\tau_{ign}$ [$\mu$s]}}
 \psfrag{x}{\small{$1000/T$[K$^{-1}$]}}
 \psfrag{p}{\small{Petersen (exp)}} 
\includegraphics[width=6.5cm]{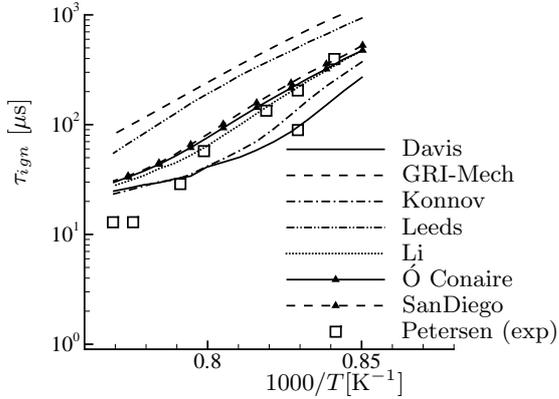} 
\caption{Ignition delay results with 2\% H$_2$, 1\% O$_2$, 97\% Ar at 33\,atm. Comparison with shock tube data of Petersen et al.~\cite{petersen1996}.}
\label{fig:petersen}
\end{figure}

Petersen et al.~\cite{petersen1996} measured the ignition delay times of stoichiometric H$_2$/O$_2$ mixtures using a high-pressure shock tube \cite{petersen1996}. The measurement data considered in the present work were performed at 33\,bar between approximately 1175\,K and 1300\,K with 2\,\% H$_2$, 1\,\% O$_2$ and 97\% dilution with Argon. Due to the relatively low temperature increase, the criteria for determining the ignition time in the computations was set to when the temperature reached 100\,K above the inlet temperature. Figure \ref{fig:petersen} show that the GRI-Mech and the Leeds mechanism largely overpredict the ignition delay data of Petersen. The latest Li mechanism involves a compromise on the equation (9a) where the low pressure limit and third body efficiencies are different depending on whether the main bath is N$_2$/He or Ar. This is done in order to take into account differences in broadening factors and temperature-dependences of collision efficiencies for different bath gases \cite{li2004}. When the ``wrong'' option is used, the Li mechanism end up on approximately the GRI-Mech and Leeds level. The Li, \'{O} Conaire and San Diego mechanisms give satisfactory agreement with the experiments in the lower temperature range and these three mechanisms also follow each other very well. At higher temperatures Konnov and Davis provide a slightly closer match with the experiments than the respective other mechanisms. The overprediction of the experiments at the highest temperatures ranges from approximately a factor two for Konnov and Davis to more than a factor 6 for GRI-mech.

\begin{figure}[tbp]
\centering
 \psfragall
 \psfrag{y}{\small{$\tau_{ign}$ [$\mu$s]}}
 \psfrag{x}{\small{$1000/T$[K$^{-1}$]}}
 \psfrag{p}{\small{H\&N (exp)}}
\includegraphics[width=6.5cm]{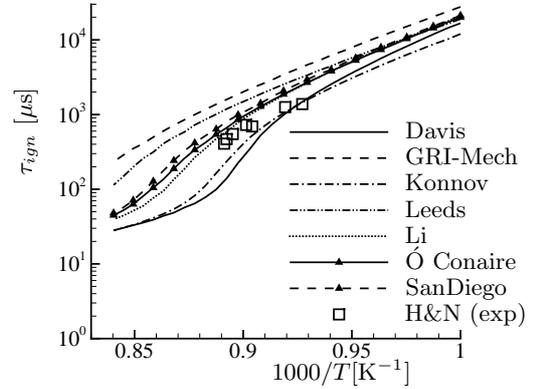} 
\caption{Ignition delay time results with H$_2$/O$_2$ diluted by 93.0\% AR at 16\,bar and $\phi$=0.5. Comparison with shock tube data of Herzler and Naumann \cite{herzler09}.}
\label{fig:herzler-phi05}
\end{figure}

\begin{figure}[tbp]
\centering
 \psfragall
 \psfrag{y}{\small{$\tau_{ign}$ [$\mu$s]}}
 \psfrag{x}{\small{$1000/T$[K$^{-1}$]}}
 \psfrag{p}{\small{H\&N (exp)}}
\includegraphics[width=6.5cm]{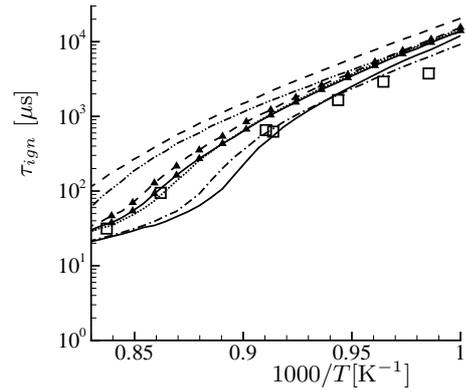} 
\caption{Ignition delay time results with H$_2$/O$_2$ diluted by 91.1\% AR at 16\,bar and $\phi$=1. Comparison with shock tube data of Herzler and Naumann \cite{herzler09}. See Fig.~\ref{fig:herzler-phi05} for legend.}
\label{fig:herzler-phi1}
\end{figure}

Recently, Herzler and Naumann \cite{herzler09} investigated the ignition of methane/ethane/hydrogen mixtures with H$_2$ content from 0\,\% to 100\,\% at 1, 4 and 16\,bar. Only the measurement data with 100\,\% H$_2$ are considered in the present work. The H$_2$/O$_2$ mixture is diluted by 93.0\% and 91.1\% Argon on mole basis at $\phi$=0.5 and $\phi$=1, respectively. The data at 16\,bar are compared with ignition delay computations at $\phi$=0.5 in Fig.~\ref{fig:herzler-phi05} and at $\phi$=1 in Fig.~\ref{fig:herzler-phi1}. The GRI-Mech and Leeds mechanisms seem to give the highest overprediction of ignition delay at all temperatures. The best agreement at low temperatures are found for the Davis and Konnov mechanisms, while Li, \'{O} Conaire and to some extend San Diego provide a closer match with the experiments at higher temperatures.

\begin{figure}[tbp]
\centering
 \psfragall
 \psfragxyign
 \psfrag{p}{\small{H\&N (exp)}}
\includegraphics[width=6.5cm]{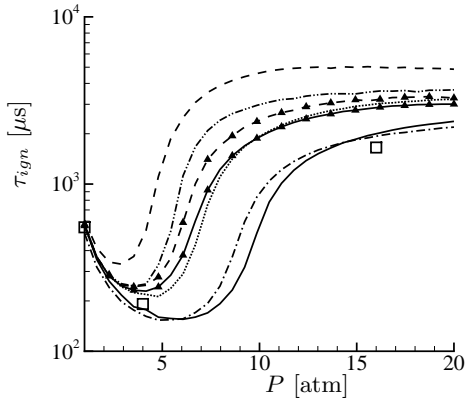}
\caption{Ignition delay time results showing the pressure effect for the composition given in Fig.~\ref{fig:herzler-phi1} at $T$=1060\,K. Comparison with shock tube data of Herzler and Naumann \cite{herzler09} at 1, 4 and 16\,bar. See Fig.~\ref{fig:herzler-phi05} for legend.}
\label{fig:herzler-peff1-phi1}
\end{figure}

\begin{figure}[tbp]
\centering
 \psfragall
 \psfragxyign
 \psfrag{p}{\small{H\&N (exp)}}
\includegraphics[width=6.5cm]{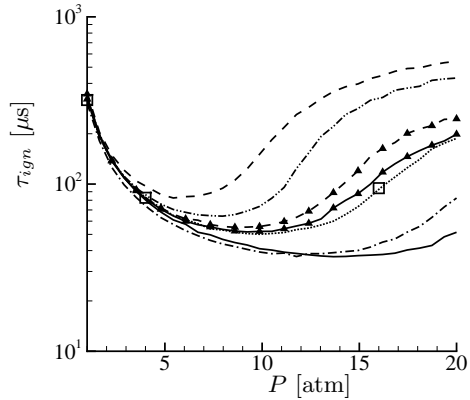}
\caption{Ignition delay time results showing the pressure effect for the composition given in Fig.~\ref{fig:herzler-phi1} at $T$=1160\,K. Comparison with shock tube data of Herzler and Naumann \cite{herzler09} at 1, 4 and 16\,bar. See Fig.~\ref{fig:herzler-phi05} for legend.}
\label{fig:herzler-peff2-phi1}
\end{figure}

Data from Herzler and Naumann are also selected to illustrate and compare the pressure effect on ignition delay time. The data are selected at approximately 1060\,K and 1160\,K\footnote{The data are selected at 1061\,K, 1064\,K, and 1059\,K at 1, 4, and 16\,bar in Fig.~\ref{fig:herzler-peff1-phi1} and at 1153\,K, 1159\,K and 1160\,K at 1, 4, and 16\,bar in Fig.~\ref{fig:herzler-peff2-phi1}.} and compared to computations in Fig.~\ref{fig:herzler-peff1-phi1} and Fig.~\ref{fig:herzler-peff2-phi1}. All mechanisms capture the trend of decreasing followed by increasing ignition delay time with increasing pressure. The shift of the minimum ignition delay time to higher pressures for higher temperatures is observed for all mechanisms. This effect is explained for the H$_2$-air cases in the following paragraph. The limited amount of data points is insufficient in order to draw any conclusions on the performance of the different reaction mechanisms. The Konnov and Davis mechanisms seem to predict the pressure effect best at 1060\,K, while the Li and \'{O} Conaire mechanisms are closer to the experiments at 1160\,K. It is worth noticing that the difference between the dedicated H$_2$ mechanisms (\'{O} Conaire, Li and Konnov) is close to a factor 5 (around 8\,bar) for the  $T$=1060\,K case and approximately a factor 2.5 (around 16\,bar) for $T$=1160\,K.

\begin{figure}[tbp]
\centering
 \psfragall
 \psfragxyign
 \includegraphics[width=6.5cm]{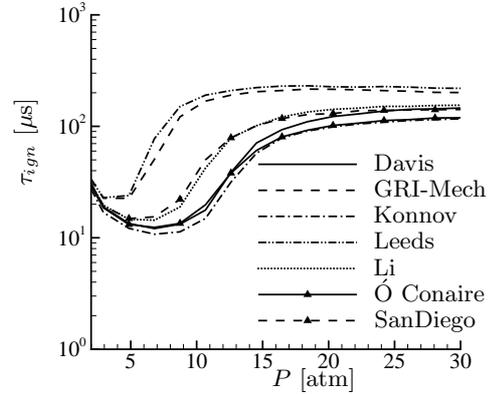} 
\caption{Ignition delay results with H$_2$-air at $\protect\Phi$=0.75 and $T$=1173\,K.}
\label{fig:h2air-T1173K}
\end{figure}

\begin{figure}[tbp]
\centering
 \psfragall
 \psfragxyign
 \includegraphics[width=6.5cm]{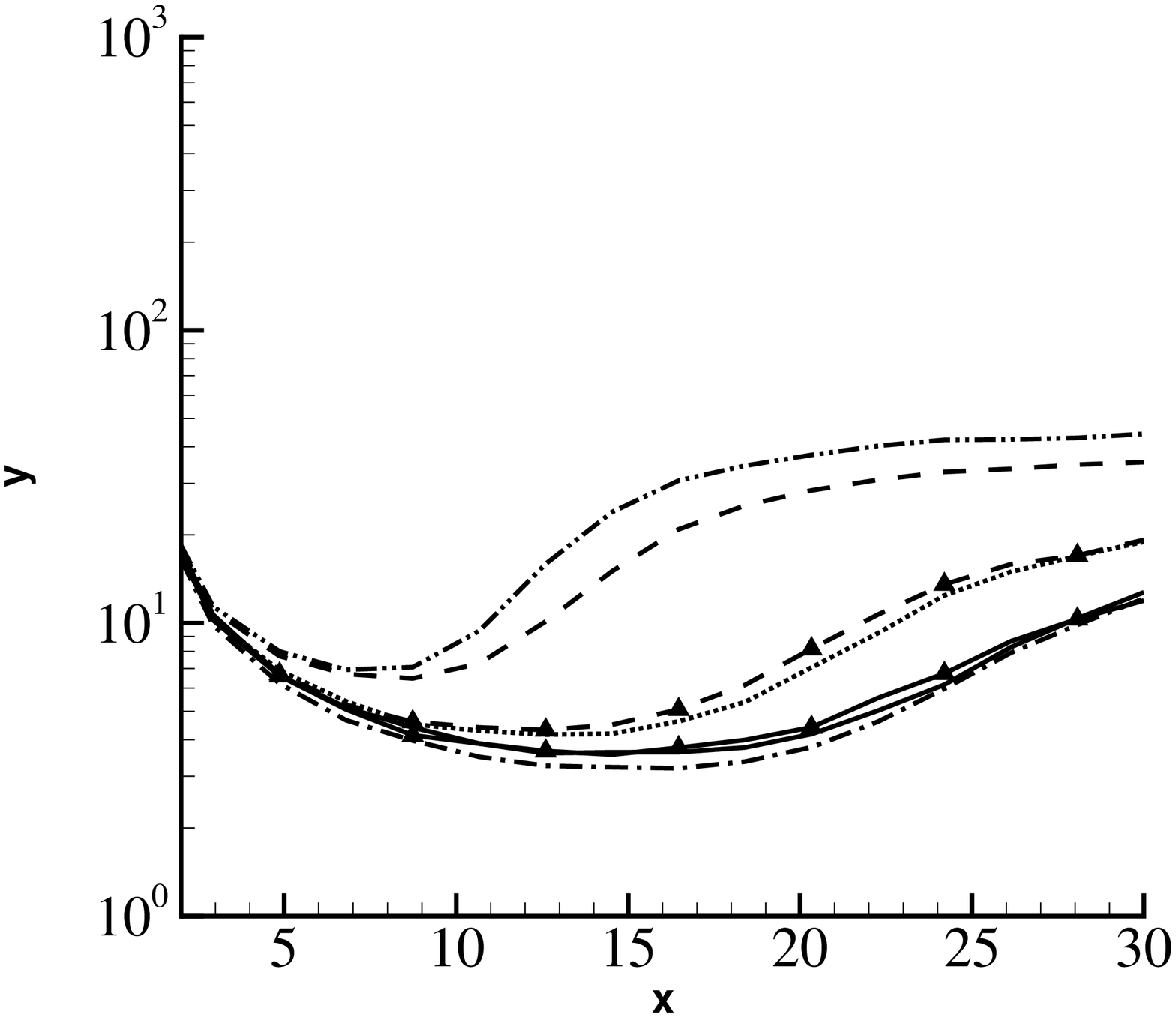} 
\caption{Ignition delay results with H$_2$-air at $\protect\Phi$=0.75 and $T$=1273\,K. See Fig.~\ref{fig:h2air-T1173K} for legend.}
\label{fig:h2air-T1273K}
\end{figure}

The remaining investigation of ignition delay time is performed with H$_2$-air at pressures and temperatures relevant to gas turbine and reheat engine conditions. No experimental work is available in the literature at these conditions so this part is limited solely to a comparison between the individual mechanisms. The Figs.~\ref{fig:h2air-T1173K} and \ref{fig:h2air-T1273K} show the ignition delay plotted against pressure for an initial H$_2$-air mixture at a fuel-air ratio of $\Phi$=$0.2$ and 1173\,K and 1273\,K, respectively. The following 4 figures cover the same temperature range at $\Phi$=$0.75$ and $\Phi$=$1.5$. The pressure trend, which is similar to the trend in the Figs.~\ref{fig:herzler-peff1-phi1} and \ref{fig:herzler-peff2-phi1}, is captured by all mechanisms. The ignition delay time drops going from 2\,bar to a minimum which lays in the range 5-20\,bar and then rises again in the interval 10-30\,bar.  A sensitivity study performed for the Li-mechanism at $\Phi$=$0.75$ and 1273\,K showed that the chain branching reaction (1) dominates at low pressures. At higher pressures the main chain terminating reaction (9) becomes increasingly important and the ignition delay time decreases. Similar sensitivity is expected for the other mechanisms. With increasing temperature, the minimum inflection point of the ignition delay is predominantly shifted to higher pressures. According to e.g.~Herzler and Naumann \cite{herzler09}, this effect is due to the activation energy of reaction (1) which increases the rate of reaction (1) relative to reaction (9) with increasing temperatures.

\begin{figure}[tbp]
\centering
 \psfragall
 \psfragxyign
 \includegraphics[width=6.5cm]{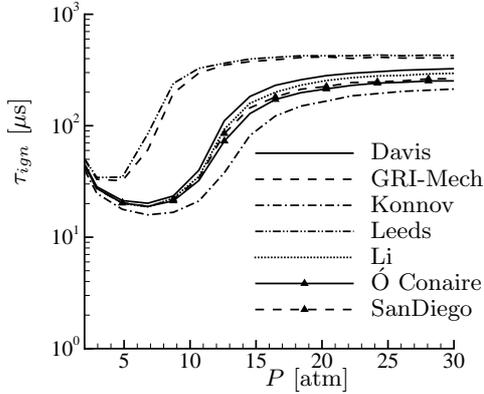} 
\caption{Ignition delay results with H$_2$-air at $\protect\Phi$=0.2 and $T$=1173\,K.}
\label{fig:h2air-T1173K-phi02}
\end{figure}

\begin{figure}[tbp]
\centering
 \psfragall
 \psfragxyign
 \includegraphics[width=6.5cm]{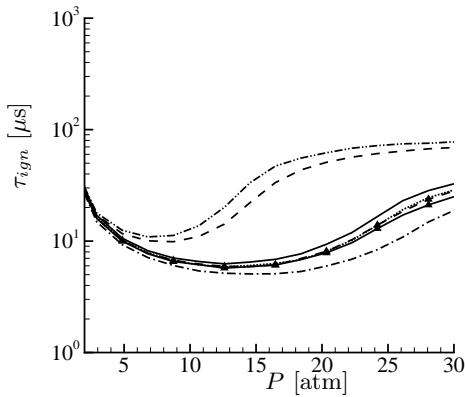} 
\caption{Ignition delay results with H$_2$-air at $\protect\Phi$=0.2 and $T$=1273\,K. See Fig.~\ref{fig:h2air-T1173K-phi02} for legend.}
\label{fig:h2air-T1273K-phi02}
\end{figure}

The discrepancies between the mechanisms are generally largest at higher pressures. This is expected since the mechanisms are better validated at lower pressures. The GRI-Mech and Leeds mechanisms predict higher ignition delay times than the other mechanisms at all temperatures, pressures and fuel-air ratios considered. The overprediction of the ignition delay time with Leeds and GRI-Mech may be attributed to the differences in the chain terminating reaction (9a). GRI-Mech and Leeds have no distinction between the high and low pressure limits, and in addition the pre-factor is significantly higher. An attempt was made where reaction (9e) of the Li mechanism was replaced by reaction (9a), (9d) and (9e) of GRI-mech for the case in Fig.~\ref{fig:h2air-T1173K}. This gave an increase in ignition delay time approaching and almost coinciding with the GRI-Mech results at higher pressures.

\begin{figure}[tbp]
\centering
 \psfragall
 \psfragxyign
 \includegraphics[width=6.5cm]{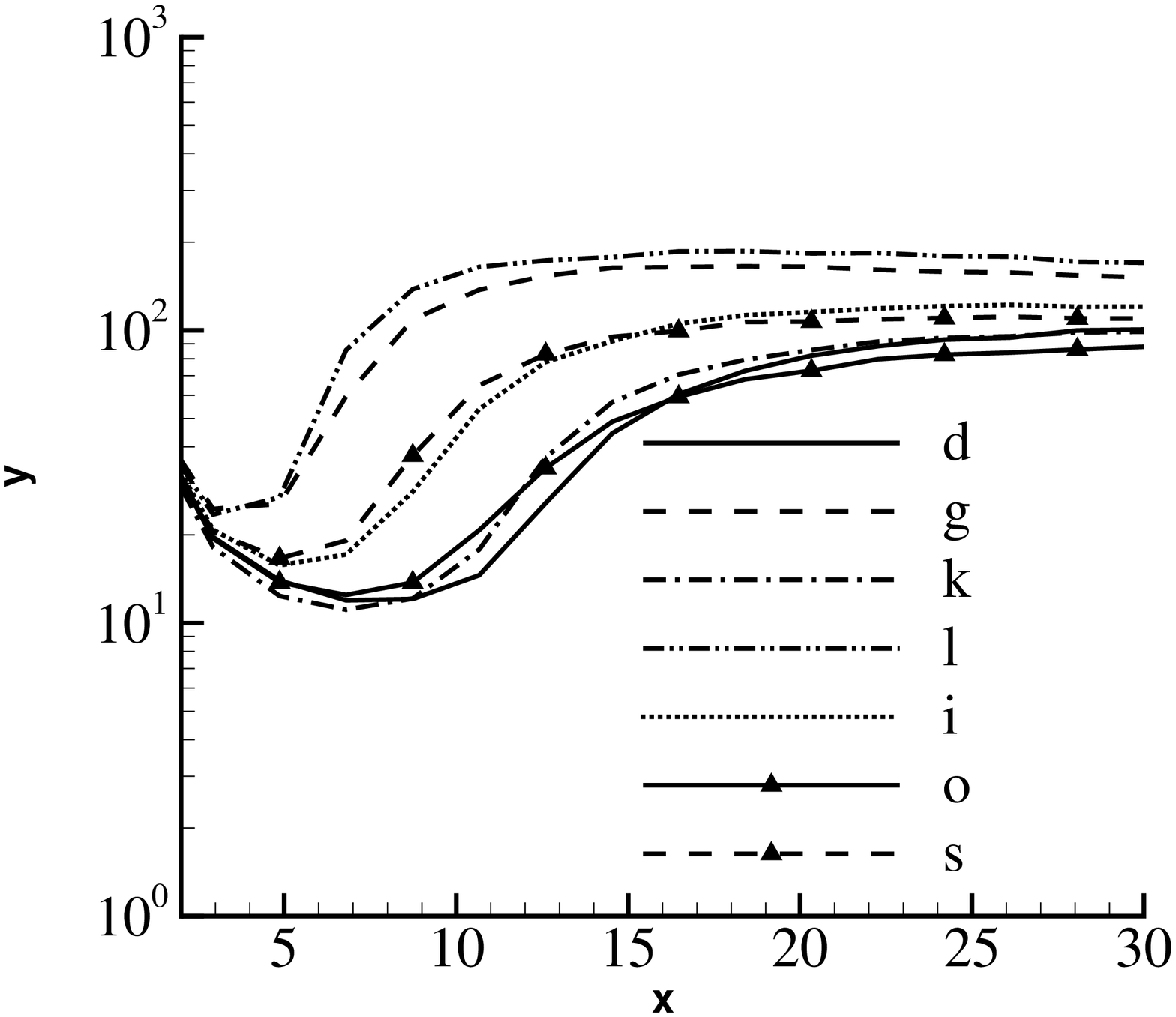} 
\caption{Ignition delay results with H$_2$-air at $\protect\Phi$=1.5 and $T$=1173\,K.}
\label{fig:h2air-T1173K-phi1p5}
\end{figure}

\begin{figure}[tbp]
\centering
 \psfragall
 \psfragxyign
 \includegraphics[width=6.5cm]{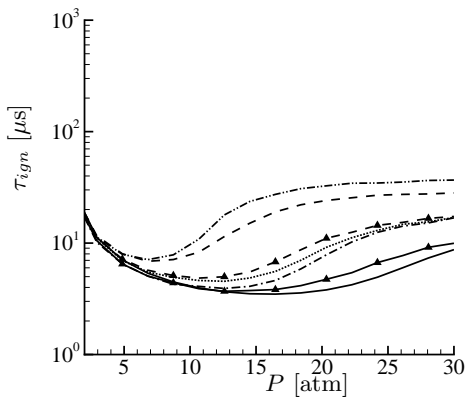} 
\caption{Ignition delay results with H$_2$-air at $\protect\Phi$=1.5 and $T$=1273\,K. See Fig.~\ref{fig:h2air-T1173K-phi1p5} for legend.}
\label{fig:h2air-T1273K-phi1p5}
\end{figure}

For $\Phi$=$0.2$ (see Figs.~\ref{fig:h2air-T1173K-phi02} and \ref{fig:h2air-T1273K-phi02}) the behavior of the mechanisms are largely divided in three groups. GRI-Mech and Leeds are significantly overpredicting the main group consisting of Davis, Li, \'{O} Conaire and San Diego, while Konnov predicts somewhat shorter ignition delay times than the main group. When $\Phi$=$0.75$ (see Figs.~\ref{fig:h2air-T1173K} and \ref{fig:h2air-T1273K}) the three groups are GRI-Mech and Leeds with the longest ignition delay times, Konnov, \'{O} Conaire and Davis with the shortest, while Li and San Diego follow each other closely somewhat above the second group. Generally, the deviation between the mechanisms increases with increasing equivalence ratio, and for $\Phi$=$1.5$ and temperatures 1273\,K (see Figs.~\ref{fig:h2air-T1173K-phi1p5} and \ref{fig:h2air-T1273K-phi1p5}), the ignition delay times are broadly distributed. It is interesting to note that, disregarding GRI-Mech and Leeds, the Davis mechanism predicts the longest ignition delay times at lean conditions, while it predicts the fastest ignition at rich conditions. This may be attributed to differences in third-body efficiencies as seen in Table~\ref{tab:reacrates}. Another interesting observation is that the Li and \'{O} Conaire mechanisms, which share the same inheritance in the Mueller mechanism \cite{mueller1999}, exhibit significant differences except at $\Phi$=$0.2$. The Li and San Diego mechanisms are very close in behavior at all conditions considered, even though the coefficients of for instance the important reactions (1) and (9a) are significantly different. The difference in ignition delay times between the dedicated hydrogen mechanisms (\'{O} Conaire, Li and Konnov) ranges from approximately a maximum factor of 1.7 at $\Phi$=$0.2$ to more than a factor 2 at $\Phi$=$1.5$. 
As a starting point in reheat burner design, the burner residence time is estimated from the ignition delay time. If the design of the reheat combustor is based on these computations, it implies that there is at least a factor 2 uncertainty margin in the size of the equipment. Taking the differences of the H$_2$/O$_2$/AR case shown in Figs.~\ref{fig:herzler-peff1-phi1} and \ref{fig:herzler-peff2-phi1}  into account, the uncertainty factor may even be closer to 5.

%
%
%
%
%

\subsection{Laminar flame speed results}
\label{sec:flame speed}
\begin{figure}[tbp]
\centering
 \psfragall
 \psfragxysl
 \includegraphics[width=6.5cm]{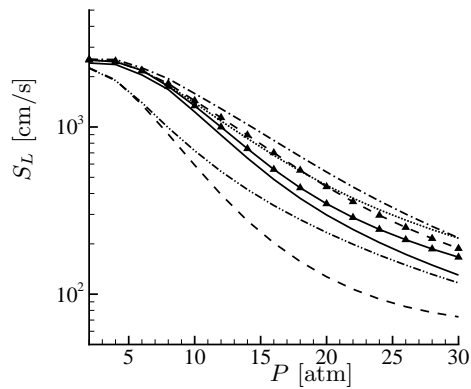} 
\caption{Laminar flame speed results with H$_2$-air at $\protect\Phi$=0.2 and $T$=1173\,K. See e.g.~Fig.~\ref{fig:SL-h2air-T1223K} for legend.}
\label{fig:SL-h2air-T1173K-phi02}
\end{figure}

\begin{figure}[tbp]
\centering
 \psfragall
 \psfragxysl
 \includegraphics[width=6.5cm]{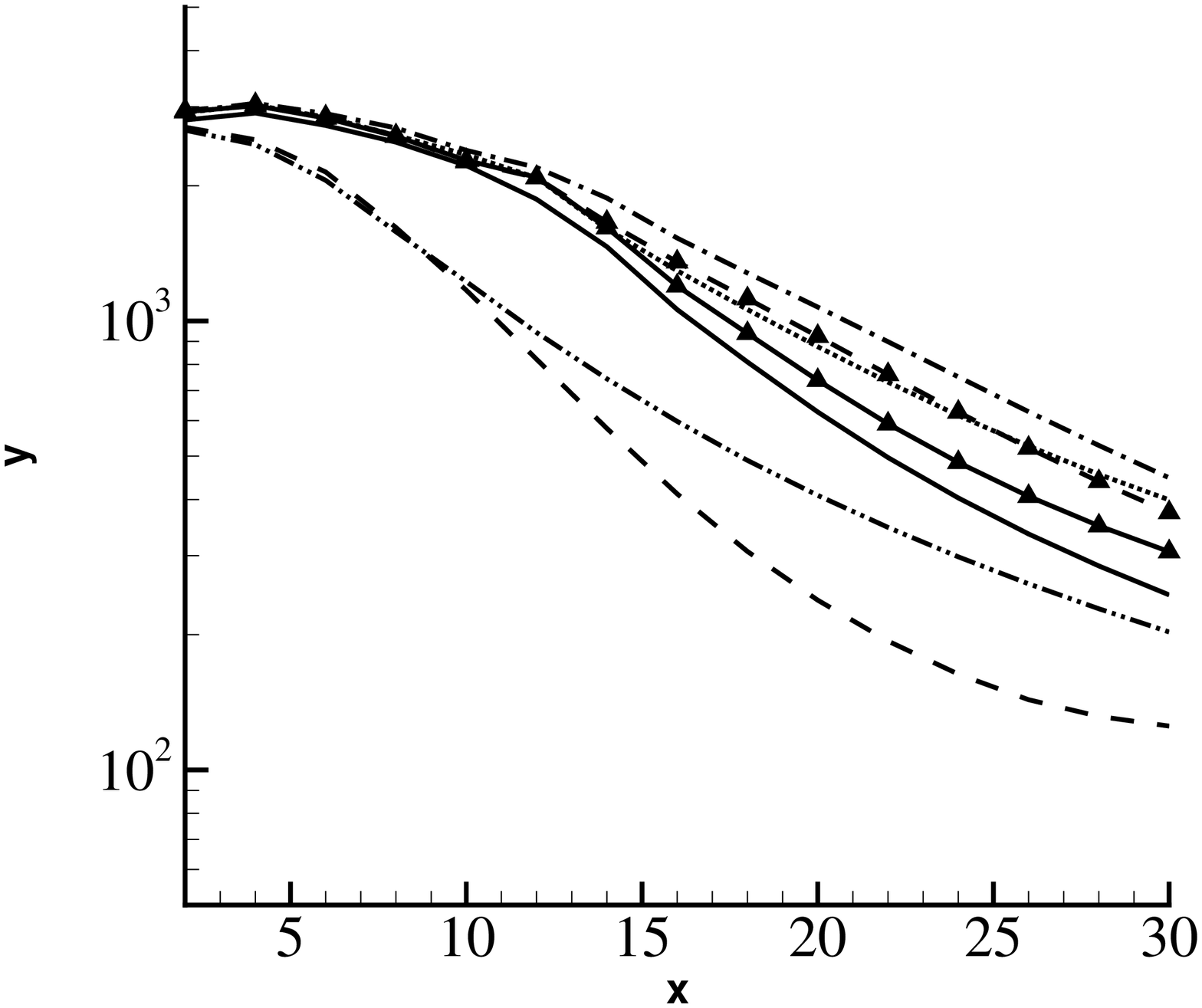} 
\caption{Laminar flame speed results with H$_2$-air at $\protect\Phi$=0.2 and $T$=1223\,K. See e.g.~Fig.~\ref{fig:SL-h2air-T1223K} for legend.}
\label{fig:SL-h2air-T1223K-phi02}
\end{figure}

\begin{figure}[tbp]
\centering
 \psfragall
 \psfragxysl
 \includegraphics[width=6.5cm]{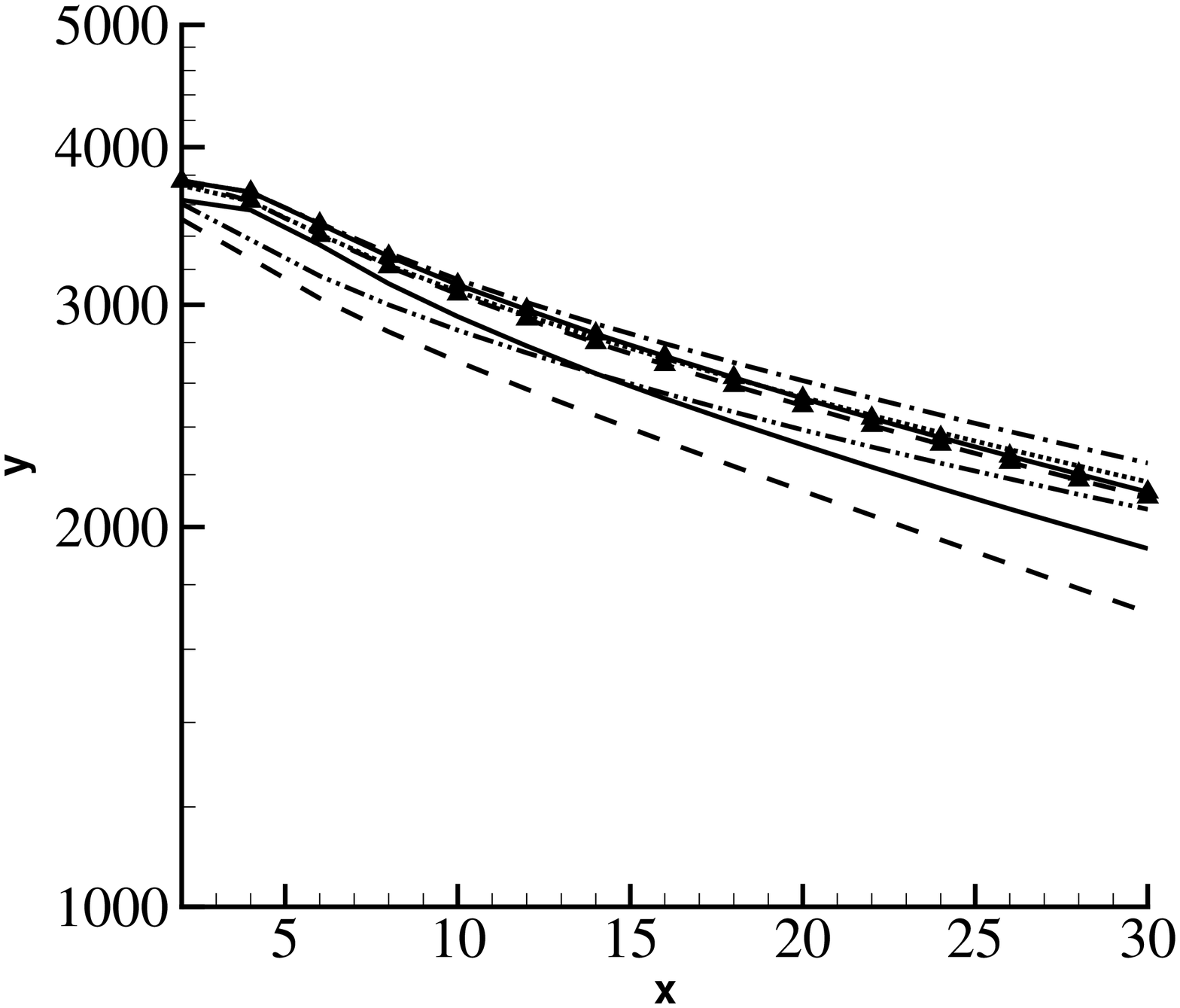} 
\caption{Laminar flame speed results with H$_2$-air at $\protect\Phi$=0.75 and $T$=1173\,K. See e.g.~Fig.~\ref{fig:SL-h2air-T1223K} for legend.}
\label{fig:SL-h2air-T1173K}
\end{figure}

\begin{figure}[tbp]
\centering
 \psfragall
 \psfragxysl
\includegraphics[width=6.5cm]{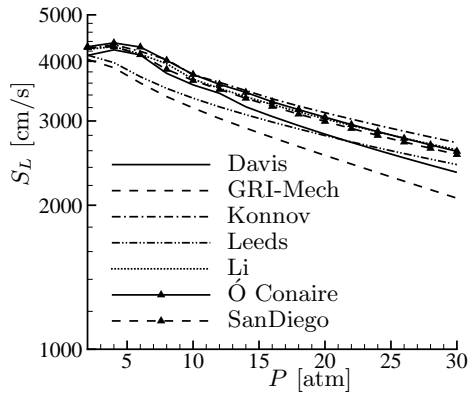} 
\caption{Laminar flame speed results with H$_2$-air at $\protect\Phi$=0.75 and $T$=1223\,K.}
\label{fig:SL-h2air-T1223K}
\end{figure}

The laminar flame speed $S_L$ computed with the PREMIX code\footnote{For laminar flame calculations with the Davis mechanism an adjusted version of the CHEMKIN transport code must be used. The code was obtained from the developers \cite{davisweb}.} are given in Figures \ref{fig:SL-h2air-T1173K-phi02} through \ref{fig:SL-h2air-T1223K} for fuel air ratios of $\Phi$=$0.2$ and $\Phi$=$0.75$, and temperatures of $T$=1173\,K and 1223\,K. At these conditions, the H$_2$-air mixture is above the auto-ignition temperature. The shortest ignition delay time for $\Phi$=$0.75$ and $T$=1223\,K is approximately 6\,$\mu$s. When the flame speed at the same conditions is about 3800\,cm/s, the distance from the upstream inlet at where auto-ignition occurs is 0.023\,cm. Considering the temperature profile in the corresponding flame speed calculations (not shown), the preheat-zone of the flame is located about 0.008\,cm from the upstream inlet. This should indicate that the flame propagation is ``ahead'' of the auto-ignition, and the results should not be significantly disturbed by this. However, the highest temperature and equivalence ratios considered in the ignition delay studies were omitted here due to problems with convergence. We also refer to the discussion about the T$_\mathrm{FIX}$ parameter in Sec.~\ref{sec:method}. 

Between approximately 5\,atm and 30\,atm the flame speed decreases with increasing pressure for all mechanisms. This is in accordance with flame speed theory, where the relative importance of the third order chain terminating reactions (9a-f) increases with increasing pressure.

The flame speed predicted by the GRI-Mech and Leeds mechanisms are generally significantly lower than the respective other mechanisms. The exception is the Davis mechanism which underpredicts the Leeds mechanism at high pressures for $\Phi$=$0.75$. At $T$=1223\,K and for both $\Phi$=$0.2$ and $\Phi$=$0.75$ the flame speed slightly increases from 2\,bar before it decreases again. This behavior is predicted by all mechanisms except the GRI-Mech and the Leeds mechanisms. A modified Li mechanism was constructed where reaction (9e) was replaced by reaction (9a), (9d) and (9e) of GRI-Mech. This modified Li mechanism did not predict the increase in flame speed at lower pressures, and approached the GRI-Mech results at higher pressures. Hence, some of the differences between the flame speed predictions with the GRI-Mech/Leeds and the other mechanisms are attributed to the differences in the chain-terminating reactions (9a-f). Here the differences between the dedicated hydrogen mechanisms (\'{O} Conaire, Li and Konnov) are less pronounced than for the ignition delay calculations. The largest deviation of about 30\,\% is found between \'{O} Conaire and Konnov at high pressures and $\Phi=0.2$.

\section{Summary and conclusions}
\label{sec:concl}
Accurate predictions of hydrogen combustion rely on high precision in the chemical mechanisms. In the present work, the performance of selected detailed hydrogen chemical mechanisms at the high temperatures and pressures relevant to gas turbine, and particularly reheat engine conditions, are investigated. For the universality of this study, pressures also higher than the baseload operating pressure of the Alstom SEV burner are considered. The effect of pressure on ignition delay time and laminar flame speed is investigated in order to illustrate the effect of variable engine load. Very little experimental data are available at these conditions so a recommended candidate among the mechanisms can presently not be selected. The GRI-Mech and Leeds mechanisms seem to give the highest overprediction of the ignition delay data at all temperatures. The best agreement at low temperatures are found for the Davis and Konnov mechanisms, while Li, \'{O} Conaire and to some extend San Diego provide a closer match with the experiments at higher temperatures.

Generally, the results with the GRI-Mech and Leeds mechanisms deviate strongly from the Davis, Li, \'{O} Conaire, Konnov and San Diego mechanisms, but there are also significant deviations between the latter five mechanisms that altogether are better adapted to hydrogen. The differences in ignition delay times between the dedicated hydrogen mechanisms (\'{O} Conaire, Li and Konnov) range from approximately a maximum factor of 2 for the H$_2$-air cases, to more than a factor 5 for the H$_2$/O$_2$/AR cases. For laminar flame speed the differences between the dedicated hydrogen mechanisms are less pronounced, with the largest deviation of about 30\,\% found between \'{O} Conaire and Konnov for H$_2$-air at high pressure. It is shown through comparison in Table \ref{tab:reacrates} and \ref{tab:additionals} that the mechanisms exhibit large differences in several important elementary reaction coefficients.

As a starting point in design, the burner residence time is chosen based on the ignition delay times.  If the predicted differences in ignition time are related to the design of actual combustion equipment, as for instance the SEV burner by Alstom, the impact on cost is significant. Hence, to reduce the uncertainties in predictions with hydrogen mechanisms at these high temperature and pressure conditions more experiments, particularly for ignition delay time, are needed. 

Generally in burner design, ignition delay and flame speed together are needed to define the characteristics. This motivates including the laminar flame speed results even at the high temperatures considered in the present work. Future work should address the role of ignition time and flame speed towards SEV flame stabilization. Finally, basic knowledge of the turbulent flame speed at such conditions is crucial for the development of hydrogen combustion engines.

\section{Acknowledgment}
\label{sec:acknowledgement}
The research leading to these results has received funding from the European Community's Seventh Framework Programme (FP7/2007-2013) under grant agreement n$^\circ$ 211971 (The DECARBit project).


\bibliographystyle{elsarticle-num}
\bibliography{h2}

\end{document}